\documentclass[aps,prb,reprint,superscriptaddress,showpacs, floatfix,longbibliography ]{revtex4-1}

\usepackage{mathrsfs}
\usepackage[cmex10]{amsmath}
\usepackage{amsfonts}
\usepackage{graphicx}
\usepackage{textcomp}
\usepackage{color}
\usepackage[caption=false]{subfig}
\usepackage{color}
\usepackage[novolumeabbr]{unitsdef}
\usepackage{esint}
\usepackage{accents}
\usepackage{bbm}
\usepackage{soul}

\renewcommand{\vec}[1]{{\boldsymbol{#1}}}
\newcommand{\op}[1]{\hat{\boldsymbol{#1}}}

\newcommand{\cross}{\times}
\newcommand{\pd}{\partial}

\newcommand{\h}{\vec{h}}
\newcommand{\e}{\vec{e}}

\newcommand{\chiz}{\op{\chi}}

\newcommand{\unitvec}[1]{{\boldsymbol{\mathbbm{#1}}}}
\newcommand{\z}{\unitvec{z}}
\newcommand{\x}{\unitvec{x}}
\newcommand{\y}{\unitvec{y}}
\newcommand{\s}{\unitvec{c}}

\newcommand{\si}{\sigma}

\newcommand{\vk}{\vec{k}_0}

\newcommand{\muz}{\mu_0}
\newcommand{\epz}{\epsilon_0}
\newcommand{\cz}{c_0}
\newcommand{\gf}{\vec{\mathfrak{f}}}
\newcommand{\gfm}{\check{\vec{\mathfrak{f}}}}
\newcommand{\ggm}{\check{\vec{\mathfrak{g}}}}
\newcommand{\gB}{\op{\mathfrak{B}}}
\newcommand{\cB}{\op{\mathcal{U}}}

\newcommand{\Q}{\op{\mathcal{Q}}}
\newcommand{\T}{\op{\mathcal{S}}}

\newcommand{\D}{\op{\mathcal{D}}}
\newcommand{\cI}{\op{\mathcal{I}}}
\newcommand{\A}{\tilde{\vec{a}}}
\newcommand{\Aj}{\tilde{\vec{a}}_{k_j}}
\newcommand{\Bj}{\tilde{\vec{b}}_{k_j}}
\newcommand{\B}{\tilde{\vec{b}}}
\newcommand{\C}{\tilde{\vec{c}}}

\renewcommand{\vr}{\vec{r}}

\newcommand{\havg}{\bar{\h}}

\newcommand{\FMR}{\text{FMR}}
\newcommand{\AFMR}{\text{AFMR}}
\newcommand{\cc}{+\text{c.c.}}
\newcommand{\iGamma}{\mathit{\Gamma}}
\newcommand{\mi}{\underline}
\newcommand{\vrho}{\vec{\rho}}
\newcommand{\vnabla}{\vec{\nabla}}

\newunit{\ohmm}{\ohm\unittimes\meter}

\DeclareMathOperator{\Imag}{Im}
\DeclareMathOperator{\Real}{Re}

\DeclareMathOperator{\sign}{sign}

\begin{document}

\title{Interaction of microwave photons with nanostructured magnetic metasurfaces}

\author{Ivan Lisenkov}
\email[]{ivan.lisenkov@phystech.edu}
\affiliation{Department of Physics, Oakland University, 2200 N. Squirrel Rd., Rochester, Michigan,
    48309-4401, USA}
\affiliation{Kotelnikov Institute of Radio-engineering and Electronics of RAS, 11-7 Mokhovaya st.,
	Moscow, 125009, Russia}

\author{Vasyl Tyberkevych}
\affiliation{Department of Physics, Oakland University, 2200 N. Squirrel Rd., Rochester, Michigan,
    48309-4401, USA}

\author{Luke Levin-Pompetski}
\affiliation{Department of Physics, Oakland University, 2200 N. Squirrel Rd., Rochester, Michigan,
    48309-4401, USA}

\author{Elena Bankowski}
\affiliation{U.S. Army TARDEC, Warren, Michigan 48397, USA}
\author{Thomas Meitzler}
\affiliation{U.S. Army TARDEC, Warren, Michigan 48397, USA}

\author{Sergey Nikitov}
\affiliation{Kotelnikov Institute of Radio-engineering and Electronics of RAS, 11-7 Mokhovaya st.,
	Moscow, 125009, Russia}
\affiliation{Moscow Institute of Physics and Technology, 9 Instituskij per., Dolgoprudny, 141700, Moscow 	Region,  Russia}
\affiliation{Saratov State University, 112 Bol'shaya Kazach'ya, Saratov, 410012, Russia}

\author{Andrei Slavin}
\affiliation{Department of Physics, Oakland University, 2200 N. Squirrel Rd., Rochester, Michigan,
    48309-4401, USA}

\begin{abstract}
A theoretical formalism for the description of the interaction of microwave photons with a thin
(compared to the photon wavelength) magnetic metasurface comprised of dipolarly interacting
nano-scale magnetic elements is developed.  A scattering matrix describing the processes of photon
transmission and reflection at the metasurface boundary is derived.  As an example of the use of the
developed formalism, it is demonstrated, that the introduction of a magnetic metasurface inside a
microstrip electromagnetic waveguide quantitatively changes the dispersion relation of the
fundamental waveguide mode, opening a non-propagation frequency band gap in the waveguide spectrum.
The frequency position and the  width of the band gap are dependent on the waveguide thickness, and
can be controlled dynamically by switching the magnetic ground state of the metasurface. For
sufficiently thin waveguides the position of the band gap is shifted from the resonance absorption
frequency of the metasurface. In such a case, the magnetic metasurface  inside a waveguide works as
an efficient reflector, as  the energy absorption in the metasurface is small, and most of the
electromagnetic energy inside the  non-propagation  band gap  is reflected.
\end{abstract}

\maketitle

\section{Introduction}
The traditional approach to the development of tunable microwave devices is to use in them  magnetic
materials magnetized externally by a variable bias magnetic field created by a combination of
permanent magnets and electromagnets~\cite{bib:Ishii:2013, bib:Helszajn:2008}.  The presence of
bulky and heavy magnets, that also bring a significant dependence of the bias magnetic field on the
temperature,  limits the applications of the magnetically biased and tunable devices in modern
microwave electronics.

On the other hand, the paradigm of reconfigurable metamaterials~\cite{bib:Zheludev:2012} and the idea of transformation optics~\cite{bib:Pendry:2006} introduced a possibility of a  precise control  of electromagnetic waves. The reconfigurable metamaterials have been demonstrated experimentally using, for example, micro-mechanical properties~\cite{bib:Lapine:2009, bib:Zheludev:2016, bib:Pryce:2010, bib:Fu:2011}, electrostatic forces~\cite{bib:Kasirga:2009, bib:Ou:2013} and temperature~\cite{bib:Tao:2009}.

However, it is highly desirable to have a reconfigurable metamaterial with ultra-short switching times,  capable of working without mechanical changes in structure and without a bias magnetic field. To address this problem a new concept of nano-structured magnetic metamaterials based on the
dipolarly coupled arrays of single-domain magnetic nanoelements has been
introduced~\cite{bib:Verba:2012,bib:Lisenkov:2015a}. The elements in these arrays are sufficiently
small to be monodomain and have sufficient  shape or crystallographic anisotropy to keep a definite
direction of their static magnetization in the absence of an external bias magnetic field. If the
anisotropy of the array element is uniaxial---each element is bi-stable, and can exist in
quasi-stable states with two opposite directions of its static magnetization. The collective static
magnetization state of an array of dipolarly coupled magnetic elements depends on the structure of the
2D periodic lattice of the array, and, also, on the magnetization ``prehistory'', and can be switched
by the application of short (less than \ilu[100]{\nanosecond}) pulses of an external bias magnetic
field~\cite{bib:Verba:2012a, bib:Verba:2013}. Obviously, when the static magnetization state of an array is
changed---the microwave absorption properties of the array are changed also, and the difference of
the microwave absorption frequencies of the same array existing in two different static
magnetization states may exceed several linewidths of the array's absorption
line~\cite{bib:Verba:2012, bib:Verba:2012a}. Between the switches the bias magnetic
field is not necessary for the functioning of the array as a passive microwave device.

The possibility to dynamically control the microwave properties of the nano-structured
magnetic metamaterials  and to use them without a permanent bias magnetic field
creates significant advantages for the devices based on these metamaterials compared to the
traditional devices based on continuous magnetic films and
multilayers~\cite{bib:Krawczyk:2014,bib:Carlotti:2014}. However, the amount of magnetic material in the
magnetic nanowire arrays is so small, that the microwave absorption in them is too small  for most
practical applications.

Therefore, the authors have proposed~\cite{bib:Lisenkov:2015} to use the arrays of coupled
magnetic nanoelements as \emph{reflectors} or \emph{metasurfaces}. In contrast with traditional
materials (e.g. ferrites) that resonantly absorb electromagnetic waves, the
metasurfaces~\cite{bib:Zouhdi:2008,bib:Yu:2014,bib:Ma:2014,bib:Rajasekharan:2015,bib:Holloway:2012, bib:Holloway:2009, bib:Kuester:2003}
significantly change the electrodynamic boundary conditions for the dynamic electric and magnetic
fields~\cite{bib:Kuester:2003, bib:Holloway:2012, bib:Yu:2014,bib:Zouhdi:2008, bib:Lisenkov:2015} near the resonant frequency of the metasurface, thus creating a strong reflection of the electromagnetic waves.

In this paper we continue to study interaction of microwave electromagnetic fields with magnetic
metasurfaces, and introduce a  scattering matrix formalism (similar to the formalism described
in~[\onlinecite{bib:Tsang:2004}]) describing the scattering of microwave photons at magnetic
metasurfaces. In the framework of this formalism the electromagnetic field is represented as a
superposition of photons with two opposite circular polarizations, and the central result of this
work is the derivation of a~photon \emph{scattering matrix} $\T$ of the nanostructured magnetic
metasurface. Having an explicit expression for the photon scattering matrix, it is straightforward
to calculate the photon transmission, reflection and/or change of spin at the interface of a
magnetic metasurface.

To illustrate the application of our formalism to the solution of a practical electrodynamic problem, we present below the calculation of the dispersion equation of the main electromagnetic waveguide mode propagating in a parallel-plate microstrip waveguide containing an array of magnetic nanowires oriented parallel to the conductive plates of the waveguide. It is important to stress, that the solution of such an electrodynamic problem is highly non-trivial, as this problem has drastically different spatial scales: the scale of the monodomain magnetic nano-element of the metasurface (nm), and the wavelength of the main electrodynamic mode of the waveguide (cm or mm). This difference in spatial scales makes the problem extremely difficult for the standard finite-difference methods. The direct numerical modeling of such a system is prohibitively time-consuming. Also, due to the fact that the dynamics of magnetization in magnetic nano-elements  comprising the magnetic metasurface is governed by the Landau-Lifshitz-Gilbert (LLG) equations, we have an additional complication related to the necessity to solve  the Maxwell equations simultaneously with  the  LLG equation~\cite{bib:Banas:2013, bib:Bruckner:2013} .

The other possible approaches to this problem include the  ``effective medium'' approach  and the multiple-scattering theory~\cite{bib:Sheng:1995}. However, a simple Maxwell-Garnett scheme can not be directly applied to the ferromagnetic elements~\cite{bib:Chang:2011}, because the magnetic permeability of a ferromagnetic element depends on the internal magnetic field, which is created by all the other ferromagnetic elements in the metasurface~\cite{bib:Gurevich:1996}.
A rigorous Clausius-Mossotti model, also, can be applied to the derivation of the  effective medium constants for a magnetic  metasurface~\cite{bib:Kuester:2003,bib:Holloway:2009}, but it requires the solution of a highly non-trivial problem of an electromagnetic wave scattering  on a nano-scale magnetic scatterer of an arbitrary shape. To escape these complications, below we propose to use a  standard spin-wave theory to  find spectra of collective spin wave excitations of a magnetic metasurface comprised of interacting magnetic elements of an arbitrary shape~\cite{bib:Verba:2012}.

We demonstrate below that using the developed formalism of the photon scattering matrix this problem can be solved analytically. In this solution we show, that the  multiple reflections of the electromagnetic wave from the magnetic metasurface substantially increase the efficiency of the interaction between the propagating wave and the metasurface. The introduction of even a very thin magnetic metasurface ($7\times10^{-4}$ times thinner than the free-space wavelength of the electromagnetic wave)  into a waveguide leads to the appearance of a non-propagation bandgap in the dispersion law of the main mode of the waveguide.  The  frequency position of the bandgap can be changed by switching the magnetic ground state of the magnetic dot array comprising the metasurface. It is also important to note, that this band gap is associated with the \emph{reflection} of the propagating waveguide mode from the magnetic metasurface, rather than with the mode absorption in this metasurface. This strong reflection is caused by the transformation of the electromagnetic field inside the waveguide caused by the necessity to fulfill the boundary
conditions for electric and magnetic fields at the upper and lower surfaces of the magnetic metasurface. We also demonstrate below, that for a sufficiently thin waveguide it is possible to choose the parameters of the metasurface and the waveguide in such a way, that the dissipation of the electromagnetic wave at the frequencies situated inside the band gap is minimized, and the waveguide containing a metasurface acts as an almost ideal reflector of electromagnetic waves.

The paper has the following structure. In Sec.~\ref{sec:photon} we derive a photon scattering matrix formalism
for magnetic metasurfaces. In Sec.~\ref{sec:waveguide} we apply the developed formalism to a problem of a wave propagation in a parallel-plate waveguide containing a magnetic metasurface. In Sec.~\ref{sec:results} we present  numerical (but not micromagnetic) results for the dispersion of a fundamental mode in a waveguide containing magnetic metasurface. The conclusions are given in  Sec.~\ref{sec:conclusions}.

\section{Interaction of photons with a magnetic metasurface}
\label{sec:photon}
\subsection{Boundary conditions}
\begin{figure}[t]
    \centering\includegraphics[width=0.8\linewidth]{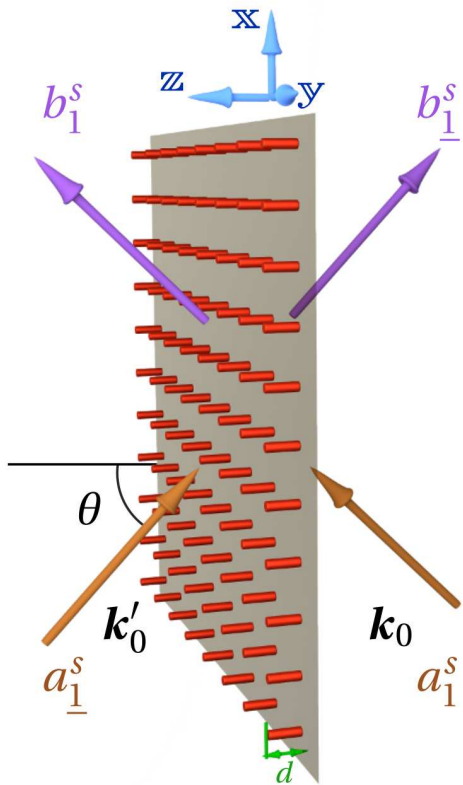}%
\caption{Picture of a magnetic metasurface exposed to electromagnetic radiation. Incident photons with
    amplitudes $a_{\pm1}^s$, wave-vector $\vec{k}$ and the incident angle $\theta$ are scattered from the metasurface and converted into the scattered photons with amplitudes $b_{\pm1}^s$. The index $s=\pm1$ defines the photon spin, $\mi1$ stands for $-1$, and $d$ is the metasurface thickness.}
\label{fig:photon_reflection}
\end{figure}

We consider a microwave electromagnetic field of the frequency $\omega$ in a free space
containing a nanostructured magnetic metasurface, see Fig~\ref{fig:photon_reflection}.
The thickness $d$ of the metasurface is assumed  to be much smaller than the wavelength $d\ll 2\pi \cz/\omega$ of the propagating waveguide mode, where $\cz$ is the speed of light.  It is also assumed, that the profile of static magnetization is uniform along the length of the nanowires comprising a metasurface.

For the further consideration we introduce an orthonormal coordinate system
$(\x,\y,\z)$. Here, $\z$ is a unit vector perpendicular the metasurface, $\x$ lays along the intersection
of the metasurface and the plane of incidence of the microwave photons (see below) , and $\y = \z\cross\x$ (see
Fig.~\ref{fig:photon_reflection}. We also assume that the metasurface is located at $z=0$.

At the metasurface the microwave electric ($\e$) and magnetic ($\h$) fields satisfy the following boundary conditions~\cite{bib:Lisenkov:2015, bib:Holloway:2012, bib:Kuester:2003, bib:Holloway:2009}:
\begin{subequations}
\begin{gather}
    \e^- - \e^+  = -i\omega\muz d\left[\z\cross(\chiz\cdot\havg)\right],\label{eq:bc_e}\\
    \h^- - \h^+ = - d (\vnabla_\vrho \otimes\z+\z\otimes\vnabla_\vrho)\cdot\chiz\cdot\havg\label{eq:bc_h},
\end{gather}%
\label{eq:bc}%
\end{subequations}
where, $\e^\pm = \e(z=\pm0)$, $\h^\pm = \h(z=\pm0)$, $\chiz$  is the external susceptibility tensor of a magnetic
metasurface~\cite{bib:Lisenkov:2015}, $\vnabla_\vrho=\x\, \pd/\pd x + \y\, \pd/\pd y$ is the in-plane
differential operator, $\otimes$ denotes the  direct vector product, and $\havg =\bigl(\h^+ +
\h^-)\bigr)/2$ is an average magnetic field acting on the metasurface.

The  electrodynamic boundary conditions  at a metasurface  that are very similar to ~\eqref{eq:bc} were used previously~\cite{bib:Holloway:2012, bib:Holloway:2009, bib:Kuester:2003} to calculate the transmission of electromagnetic waves through a metasurface using a Clausius-Mossotti-like model. The model   of a  metasurface presented in ~\cite{bib:Holloway:2012, bib:Holloway:2009, bib:Kuester:2003}    is very general, and can be applied to metasurfaces of different types.

However, the Clausius-Mossotti procedure is rather complicated technically for the metasurfaces comprised of strongly interacting magnetic elements that we are describing in our current work.  In our approach, this procedure is not necessary, because the external susceptibility tensor  chi, which we use in our boundary conditions \eqref{eq:bc}), is calculated using the spectra  of collective spin-wave excitations of the nanostructured magnetic metasurface (see Sec.~\ref{sec:results} and~[\onlinecite{bib:Lisenkov:2015}] for details). These spectra are dependent on the shapes, magnetic parameters and the lattice structure of an array of magnetic nanoelements comprising the magnetic metasurface, thus giving  a simple, but qualitatively correct description of the collective dynamic magnetic properties of the metasurface.

\subsection{Photon representation}

To solve electrodynamic problems involving magnetic metasurfaces one, typically, needs to
find a solution of Maxwell equations with the boundary conditions~\eqref{eq:bc} and other boundary
conditions defining a particular problem. The direct solution of such a system of equations in terms
of the components of vectors $\e$ and $\h$ describing dynamical electric and magnetic fields is
usually difficult, because the boundary conditions ~\eqref{eq:bc} themselves satisfy the Maxwell
equations, thus making the system of equations overdetermined and degenerate.  Of course, in each
particular case it is possible to find a projection of the equations to avoid the degeneracy, but
this difficulty has to be dealt with on a case-by-case basis.

Our way out of this difficulty will be to use a conventional scattering matrix formalism~\cite{bib:Tsang:2004}, where we operate with the \emph{complex amplitudes of photons}, which are the elementary excitations of an electromagnetic
field that satisfy the Maxwell equations. This approach simplifies the calculations considerably, and provides a general framework, that could be used to solve a variety of electrodynamic problems involving magnetic metasurfaces based on the arrays of interacting magnetic nanoelements.

First, we write  a six-dimensional electromagnetic field vector comprised of the components of the three-dimensional vectors $\e$ and $\h$ in the form:
\begin{equation}
     \gf(\vr,t) = \begin{pmatrix}\e(\vr,t) \\ \muz\cz\h(\vr,t) \end{pmatrix}.
     \label{eq:six_dim_vector}
\end{equation}
This representation looks natural, but is not convenient, since \emph{only four} components of
the vector $\gf$ are linearly independent, because the electric and magnetic fields are connected by
the Maxwell equations. Thus, below we will make several formal steps to transfer the problem from
the six-dimensional space, involving projections of the variable electric and magnetic fields,
to a four-dimensional space, involving photon amplitudes, thus removing the degeneracy of the boundary conditions~\eqref{eq:bc}.

The electromagnetic field can be represented as a superposition of photons. The photons with the
frequency $\omega$ have wavevectors $\vk$ with $|\vk|=k_0 = \omega/\cz$. Since we are interested in
the interaction of photons with a metasurface lying in the $(\x,\y)$ plane,  we consider here only the photons having equal projections of their wave vectors onto the $(\x,\y)$ plane.
Of course, there is a possibility of an alternative representation of the electromagnetic field as a superposition of the s- and p-polarized plane waves~\cite{bib:Holloway:2012}.   However, in such a case the s- and p-waves have different projections on the direction of the magnetic field at the metasurface,  and are not completely equivalent.  In contrast, when the basis of circularly polarized waves (photons) is used, the photons  having left and right circular polarizations  are absolutely equivalent.

There are four types of such photons distinguished by their direction of propagation
$\sigma=\sign(\z\cdot\vk) = \pm1$, namely propagating \emph{along} and \emph{counter} the positive direction of the axis
$\z$, and their chirality $s=\pm1$, associated with the photon spin. Without loss of generality we
can assume that the wavevector $\vk$ lies in the $(\x,\z)$ plane, which allows us to define the propagation
angle as:
\begin{equation}
\theta=\arcsin(\x\cdot\vk/k_0),
\end{equation}
and $-\pi/2<\theta<\pi/2$, see Fig.~\ref{fig:photon_reflection}.

For each of the four above introduced photon modes we can define a six-dimensional vector of the electromagnetic field $\gfm^{s}_{\sigma}$. These six-dimensional ``photon mode'' vectors will be used below as a~\emph{four-dimensional basis} in the six-dimensional space to represent the electromagnetic fields:
\begin{equation}
    \gfm^s_{\sigma} = \dfrac{1}{\sqrt{2}} \begin{pmatrix}\y \\ \s_\sigma\end{pmatrix} +\dfrac{i
        \sigma s}{\sqrt{2}} \begin{pmatrix}-
        \s_\sigma\\ \y\end{pmatrix}
    \label{eq:mode_vectors}
\end{equation}
where $\s_\sigma = -\sigma \x \cos\theta + \z \sin\theta$. Each of the photon modes carries a spin of~\cite{bib:Beth:1936}:
\begin{equation}
    \vec{s}_\sigma^s = -i\hbar \left[(\e_\sigma^s)^\dagger\cross\e_\sigma^s\right],
\end{equation}
where $\e_\sigma^s = 1/\sqrt{2}(\y - i \sigma s\, \s_\sigma) $ is a component responsible for the electric field of in the $\gfm^s_\sigma$ mode. By the definition  $|\vec{s}_\sigma^s | = \hbar$ and the direction of $\vec{s}_\sigma^s$ is collinear with $\vec{k}_0$. The  projection of
spin $\vec{s}_\sigma^s$ of each of the  photon modes on the axis $\z$ is:
\begin{equation}
    \z\cdot\vec{s}_\sigma^s= s\hbar\cos\theta,
\end{equation}
and does not depend on the direction of propagation. From the definition, it is seen, that the sign of the projection is connected with the photon chirality.

It is also convenient to introduce a dual vector basis $\ggm^s_\sigma$ to the vectors $\gfm^s_\sigma$,
the elements of which we will call~\emph{projectors} and define  it as:
\begin{equation}
    \ggm^s_\sigma = \left(\dfrac{\gfm^{s}_{\sigma} - \gfm^{-s}_{-\sigma}
        \sin^2\theta}{\cos^2\theta(3 - \cos2\theta)}\right)^\dagger.
    \label{eq:mode_projectors}
\end{equation}
One can easily check that the vectors forming the basis of the ``photon modes'' \eqref{eq:mode_vectors} and the basis of ``projectors'' \eqref{eq:mode_projectors} satisfy the following orthogonality relation:
\begin{equation}
    \ggm^s_\sigma \cdot\gfm^{s'}_{\sigma'} = \delta_{ss'}\delta_{\sigma\sigma'},
\end{equation}
where $\delta$ is the Kronecker symbol.

Using the basis of the ``photon modes'' ~\eqref{eq:mode_vectors} one can represent the dynamical
electromagnetic field as a superposition of photons traveling in the directions along and counter to the positive direction of the  $\z$ axis and having the  wavevectors
$\vk$ and $\vk'=\vk - 2\z(\z\cdot \vk)$:
\begin{multline}
    \gf(\vr,t) =\\ \sum_{s=\pm1} q^s_\sigma \gfm_\sigma^s e^{i\vk\cdot\vr - i \omega t} + \sum_{s=\pm1}
    q^s_{-\sigma} \gfm_{-\sigma}^s e^{i\vk'\cdot\vr - i \omega t} \cc
    \label{eq:photon_real_representation}
\end{multline}
where $q^s_\sigma$ are the complex amplitudes of the ``photon modes'', and $\sigma=\z\cdot\vk/|\z\cdot\vk|$. The modulus of the complex amplitude has a physical meaning of the photon density, while the argument  of this amplitude defines the phase of a particular mode.
\subsection{Scattering matrix}

The metasurface plane divides the space into two sub-spaces. In each sub-space there are
two classes of photons: the photons traveling towards and the photons traveling from the
metasurface. We shall name the photons
of the first class incident photons, while the photon of the second class scattered photons.
Fixing some vector $\vk$ and using the representation~\eqref{eq:photon_real_representation} we can express the electromagnetic fields at the both sides of the metasurface in the following form:
\begin{equation}
    \begin{gathered}
    \gf^- = \left(\sum_{s=\pm1} a^{s}_{1} \gfm^{s}_{1} + b^{s}_{\mi{1}}
        \gfm^{s}_{\mi1}\right)e^{i\vk \cdot\vrho - i\omega t}\\
    \gf^+
    = \left( \sum_{s=\pm1} a^{s}_{\mi1} \gfm^{s}_{\mi1} + b^{s}_{1} \gfm^{s}_{1}\right)e^{i\vk \cdot\vrho - i\omega t},
    \label{eq:decomp}
    \end{gathered}
\end{equation}
where $\mi{1}$ stands for $-1$ and $\rho$ is a vector lying in the $(x,y)$ plane. Here $a^s_\sigma$
is the complex amplitude of the incident photon, while $b^s_\sigma$ is the complex amplitude of the
scattered photon. Substituting these decompositions for electromagnetic fields in the boundary
conditions~\eqref{eq:bc} and regrouping terms we get:
\begin{multline}
    \sum_{s=\pm1} a^{s}_{1} \gfm^{s}_{1} - i a^{s}_{\mi1} \gB\cdot \gfm^{s}_{\mi1}
                 -a^{s}_{\mi1} \gfm^{s}_{\mi1} - i a^{s}_{1} \gB\cdot \gfm^{s}_{1}
    =\\
    \sum_{s=\pm1} b^{s}_{1} \gfm^{s}_{1} + i b^{s}_{\mi1} \gB\cdot \gfm^{s}_{\mi1}
                 - b^{s}_{\mi1} \gfm^{s}_{\mi1} + i b^{s}_{1} \gB\cdot \gfm^{s}_{1} ,
    \label{eq:bc_rewritten}
\end{multline}
where:
\begin{gather}
    \gB =- \dfrac{k_0d}{2}\begin{pmatrix}
        \op{0} & \op{M} \\
        \op{0} & \op{L}
    \end{pmatrix},\\
    \begin{gathered}
    \op{M} = (\y\otimes\x - \x\otimes\y)\cdot\chiz,\\
    \op{L} = \sin\theta  (\x\otimes\z + \z\otimes\x)\cdot\chiz,
    \end{gathered}
\end{gather}
and $\op{0}$ is the  3x3 zero matrix.

Multiplying Eq.~\eqref{eq:bc_rewritten} by the projectors $\ggm^s_\sigma$ we obtain four scalar equations, which
can be written in a matrix form as follows:
\begin{equation}
    (\cI + i\, \cB)\cdot \B = (\cI - i\, \cB)\cdot \A,
    \label{eq:bc_in_ab_form}
\end{equation}
where
\begin{equation}
    \begin{gathered}
        \A =  \begin{pmatrix} a^{1}_{1} & a^{\mi1}_{1} & a^{1}_{\mi{1}} & a^{\mi1}_{\mi1}\end{pmatrix},\\
        \B =  \begin{pmatrix} b^{1}_{1} & b^{\mi1}_{1} & b^{1}_{\mi{1}} & b^{\mi1}_{\mi1}\end{pmatrix},\\
    \end{gathered}
    \label{eq:mode_ampls}
\end{equation}
are the 4-dimensional vectors consisting of the amplitudes of the incident and scattered photons, $\cI$ is the four-dimensional identity
matrix and $\cB$ is the 4 x 4 matrix, the elements of which are calculated as follows:
\begin{equation}
    \left[\cB\right]^{ss'}_{\sigma\sigma'} = \sigma
    \ggm^{s}_{\sigma}\cdot\gB\cdot\gfm^{s'}_{\sigma'}.
    \label{eq:cB}
\end{equation}
The four-dimensional matrix $\cB$ is the projection of the six-dimensional boundary operator~$\gB$  into
the four-dimensional space, and the explicit expressions for the matrix elements of $\cB$ are presented in the Appendix.

Using these matrix elements we can, finally, write a simple expression relating the amplitudes of the scattered photons to the amplitudes of the incident photons via the scattering matrix $\T$:
\begin{equation}
    \B = \T\cdot\A,
    \label{eq:smatrix}
\end{equation}
where
\begin{equation}
    \T =  (\cI + i\, \cB)^{-1}\cdot(\cI - i\, \cB).
\end{equation}

Eq.~\eqref{eq:smatrix} is the representation of the the boundary conditions~\eqref{eq:bc} in the ``photon basis".
It is clear, that in this four-dimensional photon basis the boundary condition have a simple and compact form.
This representation of the boundary conditions at the sides of a magnetic metasurface is the central result of this
paper.  The developed formalism of the ``photon amplitudes'', similarly to the formalism of ``second quantization'' in quantum
mechanics, is coordinate-independent, making it convenient to use this formalism  in a wide class of
electrodynamic problems.  When the explicit form of the scattering matrix $\T$ is known, it is possible to solve almost any electrodynamic problem
involving a magnetic metasurface characterized by the external susceptibility tensor $\chiz$ as a standard problem in a linear scattering formalism.
We note, that a similar scattering matrix $\T$ was obtained  using the  basis  of plane  linearly polarized waves  in ~\cite{bib:Holloway:2009, bib:Holloway:2012}.

Since the linearly independent basis of our problem is four-dimensional, the symmetry properties of the 4x4 matrix $\cB$  determine all the symmetry properties of the scattering process of an electromagnetic wave from a magnetic metasurface.  For example, if the 3x3 external susceptibility tensor of a metasurface is Hermitian $\chiz = \chiz^\dagger$, the 4x4 scattering matrix $\cB$ of this metasurface is also Hermitian $\cB = \cB^\dagger$, and the scattering matrix $\T\cdot\T^\dagger=\cI$ is unitary, meaning that there is no dissipation in the process of transmission and reflection of  electromagnetic waves at this metasurface.

\section{Electromagnetic waveguide containing a magnetic metasurface}
\label{sec:waveguide}
\begin{figure}[t]
    \centering\includegraphics[width=0.9\linewidth]{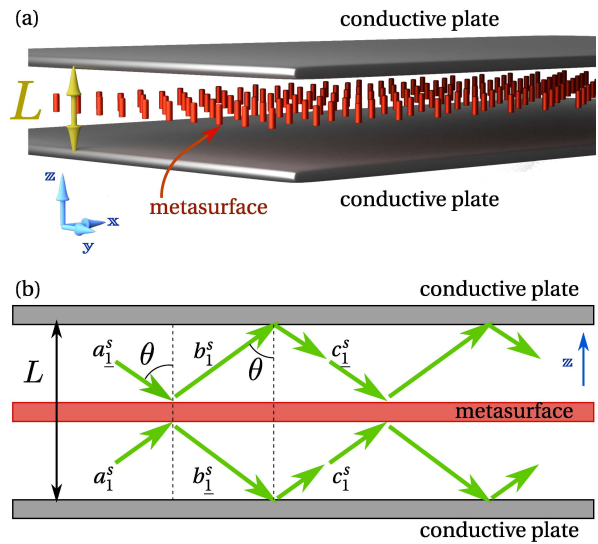}%
\caption{(a) Sketch of an electromagnetic waveguide with parallel conducting plates containing a
    magnetic metasurface situated between the waveguide plates. The magnetic metasurface is
    represented by an array of monodomain magnetic nanowires, arranged in a periodic lattice; (b) Cross-section of the
    electromagnetic waveguide. Green arrows show photon trajectories in the waveguide. Symbols
    $a^s_\sigma$, $b^s_\sigma$ and $c^s_\sigma$ stand for photon amplitudes.}
\label{fig:wg}

\end{figure}
To demonstrate an application of our theoretical formalism to a particular electrodynamic problem we
consider below the scattering of an electromagnetic wave propagating in a parallel-plate strip-line microwave waveguide of the thickness $L=2l$ from a  magnetic metasurface placed inside the waveguide at the distance $l$ from the bottom conductive plate of the waveguide, parallel to this plate (see Fig.~\ref{fig:wg}(a)). The thickness of the metasurface is $d$, and it is assumed to be small $d\ll l$.

The electromagnetic field in the waveguide must satisfy the Maxwell equations, the boundary conditions~\eqref{eq:bc} on the metasurface and the Leontovich boundary conditions~\cite{bib:Landau:1984} at the conductive plates. Instead, of facing this complex system of equations we use the developed formalism of the scattering matrices  to find the influence of the magnetic metasurface on the dispersion properties of the elextromagnetic wave propagating on a waveguide.

The electromagnetic field of any particular mode traveling in the waveguide and having the wavenumber $k$ can be represented as a set of photons~\cite{bib:Ramo:2008} reflecting between the plates with some \emph{complex} propagation angle $\theta$  with respect to the axis $\z$, see Fig.~\ref{fig:wg}(b). The photons are scattered by the metasurface, travel to the plates, than are reflected by the conductive plates, and, finally, travel
back to the metasurface. Reflection from a conductive plate reverses the photon's propagation direction $\si$ and changes its amplitude, and after the reflection from the plates the photons return to the metasurface ( see Fig.~\ref{fig:wg}(b)).  The amplitudes $c$ of these ``new'' incident photons can be related to the amplitudes $b$ of the scattered photons by the expression:
\begin{equation}
    c^s_\si = e^{2 i k_0l \cos\theta}\left[p_1 b^s_{-\si} + p_2
        b^{-s}_{-\si}\right],
\end{equation}
where $p_1 = r_1 + r_2$,  $p_2 = 1 +  r_1 - r_2$,  and the coefficients $r_1$ and $r_2$ are found
from the Leontovich boundary conditions~\cite{bib:Landau:1984}:
\begin{equation}
         r_1 = \dfrac{1}{\zeta\cos\theta - 1}, \quad r_2 = \dfrac{\zeta}{\zeta-\cos\theta},
\end{equation}
where $\zeta = (1-i)\sqrt{\omega\rho/(2\muz\cz^2)}$ is the relative impedance of the conductive surface and $\rho$ is the resistivity of the metal forming this surface.
In a stationary regime the new incident photons must be identical to the initial photons $\A = \C$. This condition leads us to the following equation:
\begin{equation}
    (\cI - \Q\cdot\T)\cdot\A = \D(\omega, \theta)\cdot\A = 0,
    \label{eq:matrix_eq}
\end{equation}
where
\begin{equation}
    \Q = e^{2 i k_0l \cos\theta}\begin{pmatrix}
        0&0&p_1& p_2\\
        0&0&p_2& p_1\\
        p_1&p_2&0&0&\\
        p_2&p_1&0&0&
    \end{pmatrix}.
    \label{eq:R}
\end{equation}
The non-trivial solutions of~\eqref{eq:matrix_eq} exist if and only if:
\begin{equation}
    \det \D(\omega, \theta) = 0.
    \label{eq:secular}
\end{equation}
This condition yields a secular equation for the waveguide modes. Finding roots $\theta_j$ of the secular
equation for a given frequency, one can obtain a dispersion relation for the $j$-th mode of a waveguide:
\begin{equation}
    k_j =  k_0 \sin\theta_j(\omega).
    \label{eq:kj}
\end{equation}
Even in the case when there is no loss of energy in the metasurface ($\chiz$ is Hermitian) and in the
conductive plates ($\rho=0$), the solution of the secular equation~\eqref{eq:secular} can be complex. The complex angle
$\theta_j$ stands for the \emph{evanescent} waves in the waveguide, and the wave number of the propagating wave in this case also becomes
complex. For the lossless case all boundary conditions are conservative, and those evanescent waves
are associated not with the damping, but with the fact, that propagating electromagnetic waves
\emph{can not} simultaneously satisfy all the boundary conditions.  This effect is, in a way, similar
to the total internal reflection in dielectrics~\cite{bib:Jackson:1975}. If the waveguide is sufficiently wide
to support several modes~\cite{bib:Ramo:2008} the secular equation ~\eqref{eq:secular} has multiple real solutions.

From the computational point of view the secular equation~\eqref{eq:secular} is an equation for a single complex variable,
and it can be solved numerically in practically all cases.

Having calculated the propagation angle $\theta_j$ for the $j$-th waveguide mode, one can substitute it back into the matrix $\D(\omega,\theta_j)$ and
calculate the vector $\Aj$ which is a non-trivial solution of this homogeneous equation. Substituting the found vector $\Aj$ for $\A$ into~\eqref{eq:smatrix} one can find the amplitudes of the scattered photons $\Bj$. Then a distribution of the electric and magnetic fields in the waveguide can be calculated from~\eqref{eq:photon_real_representation}:
\begin{multline}
   \begin{pmatrix}
       \e(\vr,t)\\
       \muz\cz\h(\vr,t)
   \end{pmatrix} = \\
   \sum_{s=\pm1} \left(a^{s}_{-\si} \gfm^{s}_{-\si} e^{-i k_0 |z|\cos\theta_j} +
                      b^{s}_{\si} \gfm^{s}_{\si} e^{i k_0 |z|\cos\theta_j}
                  \right)\times \\
                  e^{ik_0x\sin\theta_j - i\omega t} \cc,
    \label{eq:field_amp}
\end{multline}
where $\si=\sign{z}$.

Thus, we have shown, that the developed theoretical formalism of photon scattering matrices allows one to solve analytically the problem of electromagnetic wave propagation in a parallel-plate waveguide containing a magnetic metasurface and having plates of a finite conductivity. The magnetic metasurface could have an arbitrary susceptibility tensor $\chiz$, meaning an arbitrary complex magnetic ground state and an arbitrary direction of the static magnetization~\cite{bib:Verba:2012,bib:Lisenkov:2015,bib:Lisenkov:2015a}. We provided a method to compute the dispersion relation for the waveguide modes~\eqref{eq:kj} and the field distribution of each of these modes~\eqref{eq:field_amp}. It is important to note, that the developed formalism allows one to treat electrodynamic problem involving arbitrarily complex magnetic metasurfaces in a way, that is very similar to the solution of well-known problems, like  photon scattering from a conductive surface~\cite{bib:Landau:1984}.

Below, we briefly discuss the conditions of applicability of the proposed model. The boundary conditions~\eqref{eq:bc} were obtained in the magnetostatic approximation. In this approximation it is assumed, that the spin-waves in the array travel much slower than the electromagnetic waves, i.e. $v_\text{sw}\ll \cz$. For  the parameters of a typical array of magnetic nano-elements the spin-waves are rather  slow~\cite{bib:Loius:2016} $v_\text{sw}\approx1\kilometer/\second$, so this condition  is  fulfilled  naturally.  Another important assumption was made concerning the array's thickness. The external electromagnetic field acting on the array (see~[\onlinecite{bib:Lisenkov:2015}] for details) was assumed to be uniform across the array, meaning that  all the other geometric parameters of the problem should be larger than the array's thickness. This condition requires that the distance between the waveguide plates is  much larger  than the array's thickness.
These approximations considerably simplify the employed mathematical formalism. A similar problem, where some of the above limitations are relaxed, can be solved using a more rigorous approach  of Clausius-Mossotti~\cite{bib:Kuester:2003, bib:Holloway:2012},  but at a cost of much more complicated computations.

\section{Results}
\label{sec:results}

\begin{figure}[t]
\includegraphics{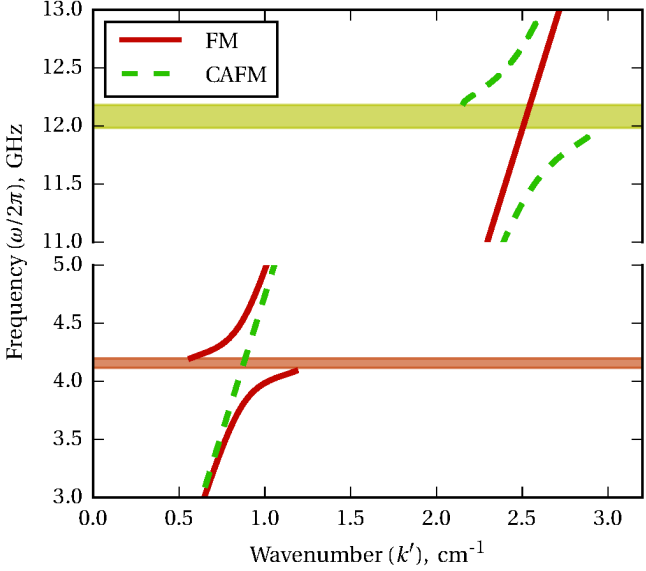}%
\caption{Dispersion relation for the fundamental mode in a flat EM waveguide containing an
    array of magnetic nanowires located in the middle between the conductive plates for the cases of
    the ferromagnetic (FM, solid line) and the chessboard antiferromagnetic (CAFM, dashed line)
    ground states of the array. The band gaps are indicated by shadowed regions. The waveguide thickness is
    \ilu[0.5]{\mm}. The parameters of the magnetic nanowire array: thickness \ilu[10]{\micrometer}, radius of a nanowire 60
    nm, the lattice constant of the square lattice 220 nm. Material properties: saturation magnetization 800 kA/m, Gilbert damping constant 0.01. Resistivity of the waveguide plates $1.68\times10^{-8}\ohmm$.}
\label{fig:disp}

\end{figure}

In our numerical example, we considered a magnetic metasurface, created by an array of magnetic
nanowires, oriented perpendicularly to the plane of the array.  The array is placed in the middle of a
parallel-plate waveguide.  The waveguide plates are assumed to be made of copper with
the electrical resistivity $\rho=1.68\times10^{-8}\ohm\unittimes\meter$. The nanowires~\cite{bib:Mel:2012} are assumed to be made of made of Permalloy, to have the height $d=10\micrometer$ and radius $r = 60\nm$, and to be arranged into a square lattice with the lattice
constant $A=220\nm$. The array can exist in two ground states, namely, the ferromagnetic (FM)state, when all the magnetic moments are orientated in the same direction, and the chessboard antiferromangetic (CAFM)state, when the nearest neighbors have their magnetic moments oriented in the opposite
directions~\cite{bib:Verba:2012}.

For these two (FM and CAFM) ground states the external susceptibility tensors are found to have the following forms:
\begin{align}
    \chiz_\text{FM} &= \dfrac{f}{2}\dfrac{\omega_M}{\omega_\FMR-\omega - i\iGamma_\text{FM}}
    \begin{pmatrix}
         1 & i & 0\\
        -i & 1 & 0\\
         0 & 0 & 0
    \end{pmatrix},
    \label{eq:chi_FM}
    \\
    \chiz_\text{CAFM} &= \zeta \dfrac{f}{2}\dfrac{\omega_M}{\omega_\AFMR-\omega - i\iGamma_\text{AFM}}
    \begin{pmatrix}
         1 & 0 & 0\\
         0 & 1 & 0\\
         0 & 0 & 0
    \end{pmatrix},
    \label{eq:chi_AFM}
\end{align}
where, for our parameters of the array, $\omega_\FMR/2\pi\approx4.06\GHz$ is the frequency of the ferromagnetic
resonance (FMR), $\omega_\AFMR/2\pi\approx12.1\GHz$ is the frequency of the antiferromagnetic
resonance (AFMR), $f=\pi r^2/A^2\approx0.23$ is the magnetic material filling fraction, $\omega_M/2\pi\approx28\GHz$
for the Permalloy, $\iGamma_\text{FM} = \alpha_G\omega_\FMR$, $\iGamma_\text{AFM} = \alpha_G\omega_\AFMR$, $\alpha_G\approx0.01$ is the Gilbert constant and $\zeta\approx1.2$ is a numerically evaluated constant, which depends on the shape of the nanowires and on the lattice
symmetry~\cite{bib:Lisenkov:2015a}. The switching between the  magnetic ground states  of  a metasurface  based on an array  of  identical magnetic nanoelements  can be done, for example, by applying short pulses of an in-plane bias magnetic field~\cite{bib:Verba:2012a}.  In the case when the array contains two types of slightly different magnetic elements  the switching can be performed quasi-statically by application of a  perpendicular magnetic field~\cite{bib:Haldar:2016,bib:Tacchi:2010}

The dispersion relation for the considered parameters of the array and the waveguide thickness
$L=0.5\mm$ is plotted in Fig.~\ref{fig:disp} for the cases of the FM (lower part of the curve) and CAFM (upper part of the curve) ground states of the
array. The thickness of the waveguide is chosen to be sufficiently small to guarantee that the cut-off
frequencies for the higher modes are larger than $\omega_\AFMR$. The dispersion relation of the
fundamental mode of the waveguide is practically unaffected by the presence of the magnetic dot
array in the frequency regions that are far from the resonance frequencies of the FM and CAFM
ground states.  At the same time, near the resonance frequencies, namely, $\omega_\FMR$ and $\omega_\AFMR$,
the dispersion relation changes drastically.  The introduction of the array opens substantial
\emph{band gaps} in the spectrum of the fundamental waveguide mode near the resonance frequencies
even in the case when the magnetic dot array is extremely thin: $\omega d/(2\pi\cz)\approx
7.1\times10^{-4}$.

\begin{figure}[t]
\includegraphics{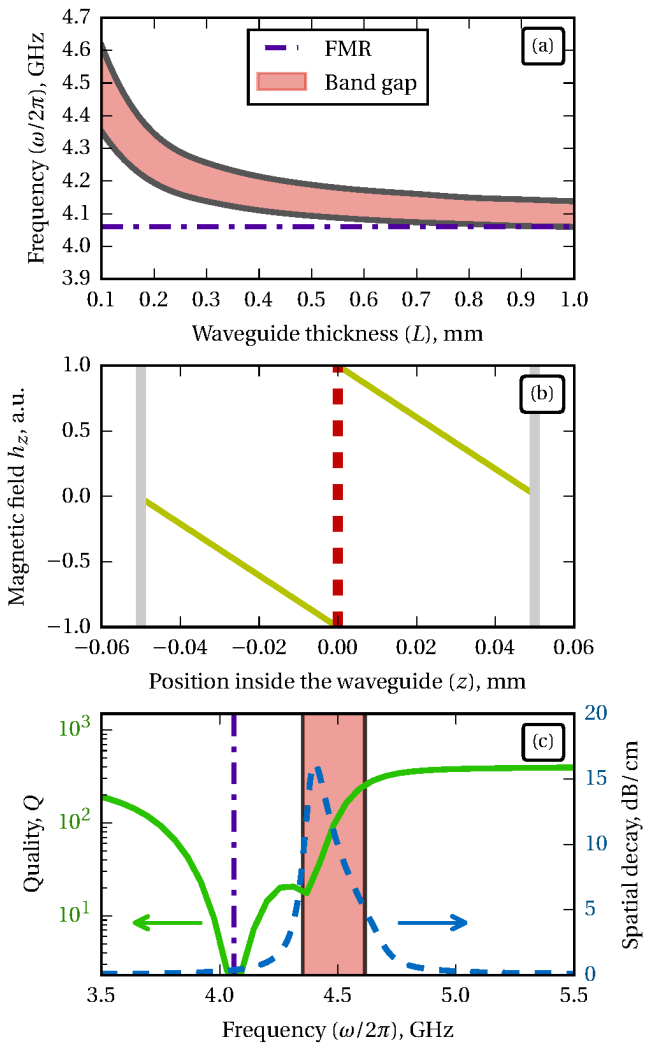}%
\caption{(a)~Dependence of the position of the band gap as a function of the thickness ($L$) of the
    waveguide for the ferromagnetic ground state; (b)~Distribution of the $\z$ component for the
    magnetic field across the waveguide for waveguide thickness $L=0.1\mm$ and frequency
    $\omega/(2\pi)=4.35\GHz$. Grey vertical lines define the positions of the conductive planes, while
    the dashed red line defines the position of the magnetic array. (c)~Quality factor (solid line, left
    axis) and spatial decay parameter of the wave along the waveguide (dashed line, right axis) versus frequency
    for the waveguide having thickness $L=0.1\mm$. The shaded area shows the position of the band
    gap. The dash-dot-dash line indicates the frequency of the ferromagnetic resonance in the both
    figures.  Parameters of the array are the same as for Fig.~\ref{fig:disp}.} \label{fig:band_gap}

\end{figure}

The band gap in the fundamental mode spectrum arises not from the losses incurred inside the array. To illustrate this we plot the dependence of the band gap width on the waveguide thickness $L$ for the FM ground state in Fig.~\ref{fig:band_gap}(a), with the dashed lines defining the frequency of the FMR. The band gap width grows with the decrease of the waveguide width, and, which is rather remarkable, the central frequency of the band gap deviates from the FMR frequency of the array for thinner waveguides.

This metasurface, having a large and almost totally reactive impedance, requires a  propagating waveguide mode to have an  in-plane  component of the electric field at the metasurface boundary to satisfy the boundary conditions~\eqref{eq:bc}. As a result of this, the propagation angle $\theta$ of the waveguide mode deviates from its ``normal'' value of $\pi/2$, the wave slows down, and the bandgap in the mode spectrum is formed.  Qualitatively, the appearance of the band gap can be understood in terms of the ``method of virtual images''~\cite{bib:Jackson:1975}. Being very good mirrors, the conductive plates of the waveguide create a virtual  ``photonic crystal'' for the photons of the main mode propagating  inside the waveguide, thus forming a band gap in its  frequency spectrum~\cite{bib:Smith:2000}.  With the decrease of the waveguide thickness, the ``reactive''  metasurface  sheet produces a progressively strong (``virtual'' metasurfaces become closer) effect and opens a larger frequency band gap (see Fig.~\ref{fig:band_gap}(a)).

The shift of the bandgap central frequency away from the FMR frequency, seen at small waveguide thicknesses, is a characteristic feature of the ferromagnetic ground state of the array, and is absent for  an array existing in the AFM state. This shift is connected with the gyrotropic properties  of the tensor $\chiz_\text{FM}$~{\eqref{eq:chi_FM}} and the boundary conditions~{\eqref{eq:bc}}, which require the presence of  non-zero $\y$ components of the electric field and a non-zero $\z$-component of the magnetic field at the location of the array. In~Fig.~{\ref{fig:band_gap}}(b) we show the distribution of the $\z$ component of the magnetic field across the waveguide for a waveguide with thickness $L=0.1\mm$. Near the conductive plates the magnetic field component is almost zero, while at the position of the metasurface (magnetic array) it is increased substantially.
Interestingly, the quasi-TEM mode has no $\y$-component in the electric field and no $\z$-component in the magnetic field, even for the propagation angle that deviates  from $\pi/2$. In terms of the waveguide modes the obtained mode for the waveguide with a metasurface in the FM state can be explained as a quasi-TEM mode coupled with one of the higher evanescent TM modes,  that have the necessary field components. The frequencies of the TM modes are higher that the frequency for the TEM mode, so  the frequency of the coupled mode is also increased,  and the band gap deviates from the frequency of the FMR.

For the frequencies lying inside the band gap, the waveguide mode becomes evanescent. This means,
that if one places a magnetic metasurface inside a waveguide and excites an electromagnetic wave outside the area where
the metasurface  is placed, this wave will be mostly reflected and some of its energy will be dissipated. The complete problem of the excitation of such a composite waveguide falls out of the scope of this paper. However, we can
estimate a \emph{quality factor} $Q$ of the waveguide containing a magnetic metasurface in the form of a magnetic nanowire
array as follows~\cite{bib:Jackson:1975,bib:Landau:1984}:
\begin{equation}
    Q(\omega) \approx \dfrac{\omega W(\omega)}{P_m(\omega) + P_e(\omega)},
\end{equation}
where $W(\omega)$ is the total stored electromagnetic energy:
\begin{equation}
    W(\omega) = \dfrac{1}{2} \int_{V}(\epz|\e(\vr)|^2 + \muz|\h(\vr)|^2)dV,
\end{equation}
$P_m(\omega)$ is the power dissipated by the magnetic metasurface~\cite{bib:Gurevich:1996,bib:Verba:2012}:
\begin{equation}
    P_m(\omega) = \omega \muz f d \int_{V} \delta(z)\Imag\left(\havg^*(\vr)\cdot\chiz\cdot\havg(\vr) \right)dV,
\end{equation}
and $P_e$ is the energy dissipated by the conductive plates~\cite{bib:Landau:1984}:
\begin{equation}
    P_e(\omega) = 2 \muz\cz \Real(\zeta) \int_{V} \delta(z-l)|\h(\vr)|^2 dV.
\end{equation}
In Fig.~\ref{fig:band_gap}(c) the frequency dependence of the quality factor is plotted for the case
of the waveguide thickness $L=0.1\mm$. The maximum absorption and the minimum of $Q$, obviously,
coincides with the frequency of the FMR. However, for such a thin waveguide the central frequency of
the band gap deviates from the frequency of the FMR,  and in the band gap region the value the
magnetic losses is much lower than at the FMR. At the same time, for the frequencies inside the band
gap the penetration depth is low, and the wave amplitude vanishes inside the waveguide very quickly
(maximum \ilu[15]{dB/cm}) on the scale of a free space wavelength (equal to\ilu[10]{\cm} in our example), see
Fig.~\ref{fig:band_gap}(c). In such a case one can expect, that the wave mode propagating in the waveguide will be mostly reflected with
practically no dissipation caused by the magnetic metasurface (nanowire array).

The variation of  the structural parameters of a magnetic dot  array on the frequencies of the FMR and AFMR  has been  studied previously~\cite{bib:Verba:2012}. In our case, this variation  shifts the position of the spectral band gap. The interaction of the incident photons with a magnetic metasurface, leading to the photon reflection,  is determined by the properties of the collective spin-wave excitations (magnons)  of  the metasurface. The magnon damping  plays a negative role in this interaction, in a sense, that the increase of damping  (characterized by the parameter $alpha_G$) leads to the  \emph{decoupling} between the magnon and photon systems, and, therefore, to the increase of the  penetration depth for the photons.  A disorder in the magnetic ground state of the array (or inhomogeneity of the array's geometrical parameters) can also lead to the additional effective damping (inhomogeneous broadening)~\cite{bib:Verba:2013}.  One possible way of  reducing the number of defects in the magnetic state  of an array by ``programming'' the element's shape  has been recently proposed  in~\cite{bib:Haldar:2016}.

In our calculations we placed the metasurface  in the middle of the waveguide  in order to make the analytical formalism  (and, in particular,~\eqref{eq:R}) simpler.  At the same time, our numerical calculations did not demonstrate  any significant influence on the metasurface position  inside the waveguide of the  dispersion of the  fundamental  mode  shown in  Fig.~\ref{fig:disp}.

\section{Conclusions}
\label{sec:conclusions}
In conclusion, we developed an analytical formalism capable of describing both qualitatively and quantitatively the interaction of electromagnetic waves (photons) with thin magnetic metasurfaces. The formalism is based on the scattering matrix method, and allows one to solve a wide variety of electrodynamic problems involving magnetic metasurfaces.

As an example of an application of our formalism we investigated the behavior of electromagnetic
waves in a parallel-plate waveguide with conducting plates containing a magnetic metasurface formed by an array of magnetic nanowires.
We found that even a rather thin magnetic metasurface introduced into the waveguide causes qualitative changes in the dispersion
of the fundamental mode of the waveguide, opening a band gap near the magnetic resonance frequency of the metasurface.
The position of the band gap depends on the magnetic ground state of the array. We showed also, that
for sufficiently thin waveguides the central frequency of the band gap deviates from the frequency
of the magnetic resonance. In this case, the waveguide can reflect electromagnetic waves with virtually no dissipation caused by the metasurface placed inside the waveguide.

\section*{Acknowledgments}
This work was supported in part by the Grant ECCS-1305586 from the National Science Foundation of the USA, by the contract from the US Army TARDEC, RDECOM, by the DARPA grant ``Coherent Information Transduction between Photons, Magnons, and Electric Charge Carriers'' and by the Center for NanoFerroic Devices (CNFD) and the Nanoelectronics Research Initiative (NRI). I.L. and S.N. acknowledge the Russian Scientific Foundation, Grant \#14-19-00760 for financial
support.

\bibliography{waveguide}

\begin{widetext}
\appendix*
\section{Matrix elements for $\cB$}
The elements of the matrix $\cB$ are found by a direct substitution of the basis vectors~\eqref{eq:mode_vectors} and projectors~\eqref{eq:mode_projectors}
into~\eqref{eq:cB}:
\begin{equation}
\cB = \dfrac{k_0 d}{8}
\begin{pmatrix}
 -u_1^a-u_2^a-u_3^s+w_{11} & i u_1^s-i u_2^s-u_3^s+w_{1\mi1} & u_1^a-i u_2^s-i u_3^a+w_{\mi1\mi1} & -i
   u_1^s-u_2^a-i u_3^a+w_{\mi11} \\
 -i u_1^s+i u_2^s-u_3^s+w_{1\mi1} & u_1^a+u_2^a-u_3^s+w_{11} & i u_1^s+u_2^a-i u_3^a+w_{\mi11} & -u_1^a+i
   u_2^s-i u_3^a+w_{\mi1\mi1} \\
 u_1^a+i u_2^s+i u_3^a+w_{\mi1\mi1} & -i u_1^s+u_2^a+i u_3^a+w_{\mi11} & -u_1^a+u_2^a+u_3^s+w_{11} & i
   u_1^s+i u_2^s+u_3^s+w_{1\mi1} \\
 i u_1^s-u_2^a+i u_3^a+w_{\mi11} & -u_1^a-i u_2^s+i u_3^a+w_{\mi1\mi1} & -i u_1^s-i u_2^s+u_3^s+w_{1\mi1} &
   u_1^a-u_2^a+u_3^s+w_{11} \\
   \end{pmatrix}
\end{equation}
where:
\begin{equation}
    w_{kn} = k\,\x\cdot\chiz\cdot\x\,\cos\theta + n\,\y\cdot\chiz\cdot\y\, \sec\theta + \z\cdot\chiz\cdot\z\,
    \sin\theta\tan\theta,
\end{equation}
and
\begin{align}
    u^s_1 &= \x\cdot(\chiz + \chiz^T)\cdot\y, && u^a_1 = i\x\cdot(\chiz - \chiz^T)\cdot\y, \\
    u^s_2 &= \y\cdot(\chiz + \chiz^T)\cdot\z\tan\theta, && u^a_2 = i\y\cdot(\chiz - \chiz^T)\cdot\z
    \tan\theta, \\
    u^s_3 &= \z\cdot(\chiz + \chiz^T)\cdot\x\sin\theta, && u^a_3 = i\z\cdot(\chiz - \chiz^T)\cdot\x
    \sin\theta.
\end{align}
Note, that if $\chiz$ is Hermitian, the coefficients $w$ and $u$ are real.
\end{widetext}
\end{document}